\documentclass[11pt]{article}

\usepackage[utf8]{inputenc}
\usepackage[T1]{fontenc}
\usepackage{xspace}
\usepackage{amsmath}
\usepackage{amssymb}
\usepackage{amsfonts}
\usepackage{hyperref}
\usepackage{fullpage}
\usepackage{url}

\usepackage{listings}
\newcommand{\Enc}{\mathsf{Enc}}
\newcommand{\MEnc}{\mathsf{MultiEnc}}
\newcommand{\Dec}{\mathsf{Dec}}
\newcommand{\pk}{\mathrm{pk}}
\newcommand{\sk}{\mathrm{sk}}

\newcommand{\intervalq}[1]{[1,#1-1]}

\newcommand{\Z}{\mathbb{Z}}
\newcommand{\F}{\mathbb{F}}

\begin{document}

\date{}
\title{Breaking the Encryption Scheme of the Moscow Internet Voting
System\footnote{This work is a merger of the papers~\cite{Pierrick19} and~\cite{Sasha19}.}}

\author{
	Pierrick Gaudry\\ 
CNRS, Inria, Université de Lorraine
	\and
	Alexander Golovnev\\
Harvard University
}

\maketitle 

\begin{abstract}
    In September 2019, voters for the election at the Parliament of the
    city of Moscow were allowed to use an Internet voting system. The
    source code of it had been made available for public testing. In this
    paper we show two successful attacks on the encryption scheme
    implemented in the voting system. Both attacks were sent to the
    developers of the system, and both issues had been fixed after that.
  
    The encryption used in this system is a variant of ElGamal over finite
    fields. In the first attack we show that the used key sizes are too
    small. We explain how to retrieve the private keys from the public
    keys in a matter of minutes with easily available resources. 
  
    When this issue had been fixed and the new system had become available for
    testing, we discovered that the new implementation was not
    semantically secure. We demonstrate how this newly found security
    vulnerability can be  used for counting the number of votes cast for
    a candidate.
\end{abstract}

\section{Introduction}

Electronic voting is more and more widely used for low-stakes elections,
with systems of various qualities. The situation for important
politically binding elections is more contrasted. Some countries have
completely banned the use of e-voting in that case (for instance, Germany
in 2009, the Netherlands in 2008, or Norway~\cite{norway} in 2013), while other
countries use it on a regular basis or organize experiments with higher
and higher stakes elections (Switzerland~\cite{chvote,swissov,scytl},
Estonia~\cite{estonia}, Canada~\cite{canada}).

The term electronic voting can cover different situations, and in this
work, we are interested in Internet voting, not machine-assisted voting
that takes place in polling stations. This increases the difficulty to
guarantee properties like authentication or coercion-resistance that are
easier to obtain at a polling station, where an officer can check
classical identity cards and where the voters can go to a polling booth to
isolate themselves and choose freely.

But even more basic properties like vote secrecy and verifiability are
not easy to obtain if one wants to keep things simple and without
advanced cryptographic tools like zero-knowledge proofs, proof of
equivalence of plaintexts, oblivious transfer, etc.

For high-stakes elections, a bad practice that tends to become
less accepted by the population is to have some designated experts that
study the security of the product, but how it really works remains secret
to voters. Therefore, in more and more cases, the organization will ask
for a product that can be audited by independent experts, and as an
incentive to have more feedback, public testing with an associated bug
bounty program can be organized.  For instance, this has been recently
the case in Switzerland, which is a country with a long history of
experiments with Internet voting. A security problem was actually
discovered at this occasion~\cite{swisspostattack1,swisspostattack2}.
\smallskip

In Russia, September 8, 2019 was a day of local elections, where
governors and representatives for local parliaments must be elected.  In
Moscow, at the occasion of this election for the City Parliament (Moscow
Duma), it was decided to test the use of Internet voting. Voters from 3
electoral districts (among a total of 45 districts) were allowed to
register for using Internet voting instead of using classical paper
voting at polling stations.

The voting system used for this election was designed specifically. For
lack of a proper name, we will call it the Moscow Internet voting system.
Its deployment is the responsibility of a service of the City called the
Department of Information Technology. In July, the system was opened for
public testing.

\subsection*{Description of the public challenge}

On July 17, 2019, some of the system's code was posted
online~\cite{github}, and the organizers asked the public to test several
attack scenarios~\cite{testing}. A bounty program of up to 2 millions
rubles (approx. \$30,000) was associated to it.  We believed that the fact that most of the
information is in Russian and that almost no description of the system
(in any language) is available apart from the source code was a reason for
having a low advertisement of this challenge at the international level,
even among the e-voting community.

The system is poorly documented, but from the source code and brief
descriptions of the system~\cite{medium}, we know that it uses the
Ethereum blockchain~\cite{ethereum} and ElGamal encryption. No advanced cryptographic
tools are present in the source code (no verifiable mixnets~\cite{mixnet}, for
instance, while they are quite frequent in modern systems).

In one of the attack scenarios, the organizers publish a challenge
consisting of the public key and some encrypted messages. The attack was
considered successful if the messages got decrypted within 12 hours (the
duration of the future, real election), before the organizers reveal the
private key and the original messages.  All of these cryptographic
challenges (keys and encrypted data) were put in the public repository
of the source code, in a special sub-directory called {\tt
encryption-keys}.

\subsection*{Contributions}

In this paper, we describe two attacks that we mounted on the system,
following this attack scenario. The
first attack uses the fact that the key sizes are so small that, with
specialized software, it is possible to compute discrete logarithms and
deduce the private keys in far less than the 12 hours allowed for this
task. After this, the source code was modified. Our second attack is
against this new version and relies on 
a subgroup attack that reveals one bit of information related
to the original message. In an e-voting context, this can be enough to
get a lot of information about the voter's choice, and indeed, in the Moscow
system, the leakage was really strong. During August, several public
tests were done, with volunteers, after the system was patched against
our attacks.  In this work, after describing our attacks, 
we will discuss the general
protocol, which is some kind of moving target, since there is no proper
specification, no clear security claims and on top of that, deep changes
were made until very late before the real election.
\medskip

For this work, we used the following different sources of information
about the Moscow Internet voting system:
\begin{itemize}
    \item The public source code, of course. This includes Javascript code
        to be run on
        the client side, PHP code for the server side, and Solidity code
        to be run as smart contracts in an Ethereum blockchain.
    \item The articles published in the press, sometimes quoting the
        designers of the system. This includes various sources, with
        different opinions about the use of Internet voting in this
        context. We considered some of these sources as non-reliable.
    \item Private discussions with the designers and with journalists
        investigating the current situation.
\end{itemize}
In the following, we will refer to  different versions of the source
code. In order to make our terminology precise, we give the exact
revision numbers of these versions, corresponding to {\tt git} commits in
the public repository~\cite{github}:
\begin{itemize}
    \item The ``original'' version, i.e. the one that was published and
        used for the first public test: revision d70986b2c4da.
    \item The ``modified'' version, that took into account our first attack:
        revision 1d4f348681e9.
    \item The ``final'' version that was used for the election: revision
        51aa4300aceb.
\end{itemize}

\section{Attacks on the encryption scheme}

\subsection{Attack on the original implementation}

In the original version of the source code (rev d70986b2c4da), the encryption
scheme can be found in the files {\tt elGamal.js} and {\tt
multiLevelEncryptor.js} of the {\tt
smart-contracts/packages/crypto-lib/src/} subdirectory.
The first file contains a textbook version of the ElGamal encryption
algorithm, while the second one builds on top of it a ``multilevel''
variant that we are going to describe here since this is a non-standard
construction.

Let us first fix the notations for the textbook ElGamal encryption. Let
$G$ be a cyclic group generated by $g$ of order $q$. An ElGamal keypair
is obtained by choosing a (secret) decryption key $\sk$ as a random
integer in $\Z_q$, and the corresponding (public) encryption key $\pk$
is given by $\pk = g^\sk$.
Let us denote by $\Enc_{g,\pk}(m) = (a, b)$ the ElGamal encryption of the
message $m\in G$ with a public key $\pk$ and a generator $g$. This is a
randomized encryption: an integer $r$ is picked uniformly at random in
$\Z_q$, and then the encryption is obtained as
$$ \Enc_{g,\pk}(m) = (a, b) = \left(g^r, \pk^r\cdot m\right).$$
The corresponding decryption function $\Dec_{g,\sk}(a,b)$, that uses the
secret key $\sk$ corresponding to $\pk$ is then given by
$$ \Dec_{g, \sk}(a,b) = b\cdot a^{-\sk} = m.$$

The multilevel variant is obtained by successively applying the ElGamal
encryption, with three different parameter sets, first on the message
$m$, and then on the $a$-part of the successive ElGamal ciphertexts. In
the Moscow system, there are 3 levels. Each level uses a group $G_i$
which is the multiplicative group of a finite field $\F_{p_i}$, where
$p_i$ is a safe prime.
An important remark, here, is that the
$p_i$'s being different, there is no algebraic map from one group to the other.
It is necessary to lift an element of $\F_{p_1}^*$ to an integer in
$\intervalq{p_1}$ before mapping it to $\F_{p_2}^*$. This mapping will be
without loss of information only if $p_2$ is larger than $p_1$; and
similarly we need $p_3$ bigger than $p_2$. These conditions are indeed
enforced in the source code.

Let us denote by $g_1$, $g_2$, $g_3$ the generators of the 3 groups
$G_1$, $G_2$, $G_3$. There are 3 ElGamal key pairs $(\sk_1,\pk_1)$,
$(\sk_2,\pk_2)$, $(\sk_3,\pk_3)$ used for the encryption and decryption of
the ballots. In order to encrypt a message $m\in G_1$, we compute the
following successive ElGamal encryptions:
$$
\begin{array}{rcl}
    ( a_1, b_1 ) & := & \Enc_{g_1,\pk_1}(m); \quad \text{map $a_1$ to $G_2$;} \\
    ( a_2, b_2 ) & := & \Enc_{g_2,\pk_2}(a_1);\quad\text{map $a_2$ to $G_3$;} \\
    ( a_3, b_3 ) & := & \Enc_{g_3,\pk_3}(a_2),\\
\end{array}
$$
and then the ciphertext is the quadruple in $G_1\times G_2 \times G_3^2$
given by
$$ \MEnc(m)= (b_1, b_2, a_3, b_3).$$

The values $a_1$ and $a_2$ are forgotten, but someone knowing the private
keys $\sk_1$, $\sk_2$, $\sk_3$ corresponding to $\pk_1$, $\pk_2$,
$\pk_3$, will be able to recover $m$ from the ciphertext with the following
decryption procedure:
$$
\begin{array}{rcl}
    a_2 & := & \Dec_{g_3,\sk_3}(a_3, b_3); \quad \text{map $a_2$ to $G_2$;} \\
    a_1 & := & \Dec_{g_2,\sk_2}(a_2, b_2); \quad \text{map $a_1$ to $G_1$;} \\
    m   & := & \Dec_{g_1,\sk_1}(a_1, b_1).\\
\end{array}
$$

The purpose of this multilevel encryption is not known to us. We will
speculate on this in Section~\ref{sec:discuss}. An obvious observation,
however, is that if the discrete logarithm problem is not hard in $G_1$,
$G_2$ and $G_3$, then it is possible to deduce the secret keys $\sk_i$'s
from the public keys $\pk_i$'s and an attacker can then decrypt encrypted
messages as quickly as the legitimate possessor of the secret keys.

In the published source code, the primes $p_i$'s have less than 256 bits.
Discrete logarithms in finite fields defined by such small primes have
been computed for the first time in the middle of the 90's: Weber, Denny
and Zayer did a series of computation in 1995-1996, starting from 215 to
281 bits~\cite{WDZ9596}.
At that time, the computing resources required for the computations were
rather high, and solving the 3 discrete logarithm problems to get the
private keys would not have been easily feasible in less than 12 hours as
required by the challenge.

More than 2 decades later, computers are much faster and have much more
memory. Furthermore, the Number Field Sieve algorithm~\cite{LeLe93},
which is the
fastest known method asymptotically was still a very
new algorithm in the mid-90's, and many
theoretical and practical
optimizations have been developed since
then~\cite{Schirokauer93,JoLe03,NFSCfrac,fastbw}.
The current record is a computation modulo a 768-bit prime~\cite{dl768}. 

We have tried the following software products that contain a full implementation
of discrete logarithm computations in prime fields:
\begin{center}
\begin{tabular}{|c||c|c|c|}
    \hline
    Software & SageMath~\cite{sagemath} & Magma~\cite{magma} & CADO-NFS~\cite{cadonfs} \\
    \hline
    Version  &  8.8     & 2.24-2 & rev. 6b3746a2e \\
    \hline
\end{tabular}
\end{center}
Note that Magma is proprietary software, while the others are free
software.

The experiments were first made on a typical personal computer equipped
with a 4-core Intel i5-4590 processor at 3.3 GHz and 16 GB of RAM.  It is
running a standard Debian distribution.  SageMath uses
GP/Pari~\cite{gppari} internally
for computing discrete logarithms. On this machine, the computation took
more than 12 hours, and actually we had to stop it after 4 days while it
was still running.
According to GP/Pari documentation, the algorithm used is a linear sieve
index calculus method. As for Magma, the handbook tells us that depending
on arithmetic properties of the prime, the algorithm used can be the
Gaussian integer sieve or a fallback linear sieve. The prime we tested
was compatible with the Gaussian integer sieve. But during the linear
algebra step, the memory requirement was much larger than the available
16 GB. We started the computation again, on a 64-core server node with 192
GB of RAM. On this machine, Magma computed the discrete logarithm in a
bit less than 24 hours with 130 GB of peak memory usage. It should be
noted that both Magma and SageMath use only one of the available
computing cores, so that there does not seem to be an easy way to go
below the 12 hours limit with them, even with an access to a powerful
machine.

CADO-NFS is an implementation of the Number Field Sieve for integer
factorization and discrete logarithms in prime fields (and some
experimental support for small degree extensions of prime fields).
The last stable release 2.3.0 is two years old, so we used the
development version, available on the public git repository.
With CADO-NFS, on the standard personal machine, the running times to
retrieve the private keys of August 18 were as follows:
\begin{center}
\begin{tabular}{|c|c|}
\hline
     key number &   time   \\
\hline
        1       &  425 sec \\
        2       &  507 sec \\
        3       &  314 sec \\
\hline
\end{tabular}
\end{center}
Note that the variation in the running time from one key to the other is
not unusual for computations with moderately small primes. Also, we
should mention that when doing this work, we realized that the development
version was not robust for numbers of
this size: it sometimes failed in the final step called ``individual
logarithm'' or ``descent''. The revision number we gave above corresponds
to a version where we have fixed these problems, so that CADO-NFS can
reliably compute discrete logarithms in finite fields of about 256 bits.

CADO-NFS
does not include the ``front-end computation'' for the discrete
logarithm: the small Pohlig-Hellman step due to the fact that the order of the
generator is twice a prime must be done by hand; similarly, the base for
the discrete logarithm computed by CADO-NFS is arbitrary. Therefore, in
order to compute one of the $\sk_i$, the program must be run twice, once
for the generator and once for the public key.  Fortunately, in the
Number Field Sieve algorithm, many parts of the computation can be shared
between the two executions modulo the same prime (this is the basis of
the LogJam attack~\cite{logjam}), and CADO-NFS indeed
shares them automatically. The running times given above include those 2
runs for each key.
For completeness and reproducibility, we
provide in Appendix~\ref{app:dlog} a script to obtain the keys; this includes the
few additional modular operations to be done apart from the calls to
CADO-NFS.
\medskip

Of course, for a real attack, the three private keys can be computed
simultaneously on 3 machines in parallel. Indeed, the chaining involved
in the multilevel ElGamal is not relevant for the keys, it occurs only
during the encryption / decryption of messages.
 
Additionally to this immediate 3-fold parallelism for the attack,
CADO-NFS also has some parallelism capabilities so that machines with
more cores can reduce the time for a single key.  However, there is some
limit to it with the current implementation. For instance, the private
key number 1 still required 160 seconds of wall clock time on the same
64-core machine that we used for testing Magma. 

\subsection{Attack on the modified version}
\label{sec:second_attack}
After the first attack was sent to the developers of the system and made
public a few days later, the public source code has been modified. The
key size has been increased to 1024 bits, and the multilevel ElGamal has
been removed and replaced by a single ElGamal encryption.

In the original version, the generators in all the involved groups were
generators for the full multiplicative group of the finite fields, thus
their orders were twice a prime numbers. This exposed the danger of
leakage of one bit of information on the message, with a subgroup attack.
This is an old technique~\cite{oldsubgroup}, but there are still frequent attacks,
in particular when an implementation forgets the key validation
step~\cite{measuringsubgroup}.
Although we did not push in this direction in the first attack, it was
explicitly mentioned as a weakness. Therefore in the modified version,
the generator was chosen to be a quadratic residue, thus having prime
order.

We discovered however that the other parts of the implementation were not
changed accordingly, so that an attack was still possible.

Let $p=2q+1$ be the 1024-bit safe prime used to define the
group, where $q$ is also a prime. Let $Q_p$ be the group of quadratic
residues modulo $p$; it has order $|Q_p|=(p-1)/2=q$. The chosen generator
$g$ belongs to $Q_p$, and therefore, so is the public key $\pk$, since it
is computed as before as $\pk=g^\sk$, where $\sk$ is randomly chosen in
$\Z_q$.

The problem with the modified implementation is that the message $m$ is
allowed to be any integer from $\intervalq{q}$ which is naturally
mapped to an element of $\F_p^*$. For semantic security (under the
Decisional Diffie-Hellman assumption), the message $m$ should instead be
encoded as one of the $q$ elements of the group $Q_p$ generated by $g$.
In the case where $m$ is not necessarily picked from the group of
quadratic residues, the Decisional Diffie-Hellman assumption does not
hold and indeed it is possible to build an efficient distinguisher, thus
showing that the encryption scheme in the modified version 
is not semantically secure.

Let us make this explicit.
If the message $m$ becomes a quadratic residue after being mapped to
$\F_p^*$, then for \emph{every} choice of randomness of the encryption
algorithm, in the resulting ciphertext $\Enc_{g, \pk}(m)=(a, b)$,
the second component 
$b$ is also a quadratic residue.  Indeed, if $g$ and $m$ belong to $Q_p$,
then there exist $x$ and $y$ in $\F_p^*$ such that $g=x^2$ and $m=y^2$
Then
$$
  b=\pk^r\cdot m = g^{r\cdot \sk}\cdot y^2=(x^{r\cdot \sk}y)^2 \in Q_p \, .
$$
Similarly, if $m$ is not a quadratic residue, then $b = \pk^r\cdot m$ is not
a quadratic residue either.

Testing the quadratic residuosity of $b$ can be done by computing the
Legendre symbol of $b$ and $p$. Thanks to the law of quadratic reciprocity, a
very efficient algorithm similar to the Euclidean algorithm is
available~\cite{GatGer}.
Therefore from just the knowledge of a ciphertext, it is possible to
immediately deduce if the corresponding cleartext $m$ belongs or not to
$Q_p$. Roughly half of the messages are mapped to $Q_p$. Hence, one bit
of information is leaked.

In order to test the validity of this attack, we checked whether the
$b$-parts of the published encrypted messages belonged or not to $Q_p$.
It turned out that exactly five out of the ten were 
quadratic residues modulo $p$. This shows that indeed, some of the
cleartexts were in $Q_p$ and some were not. Details for reproducing
these computations are given in Appendix~\ref{app:legendre}.

\subsection{On the role of encryption in the protocol -- What did we break?}

As in many e-voting protocols, the encryption scheme is used to encrypt
the choice of the voter to form an encrypted ballot. From the Javascript
source code (under a sub-directory called {\tt voting-form}) that is
supposed to be run on the voting device of the voter, we deduce that the
encrypted data consists solely of this choice (with no additional nonce or
meta-data). It takes the form of a 32-bit unsigned integer called
``deputy id'' that looks random.

The link between the deputy ids and the real names of the candidates is
public, since the Javascript source code that must present the choices to
the voters has to include it.

In the original version of the encryption scheme, as soon as the election
starts, the 3 public keys of the multilevel ElGamal must become public,
and from them, in a matter of minutes the decryption keys can be deduced.
Then, this is as if the choices of the voters were in cleartext all along
the process. Even if there is a strong trust assumption on the server
that receives these votes, and even if it is honest and forgets the link
between the voters and the ballots, there is still the issue of putting
them in the blockchain for verifiability. Since the ballots are
(essentially) in cleartext, the partial results become public all along
the day of the election, which can have a strong influence on the result.
Actually, it is illegal in Russia to announce any preliminary result
while the election is still running.

Our second attack will not give a full information. Just one bit of
information is leaked from an encrypted ballot, namely whether or not the
chosen candidate has a deputy id which is a quadratic residue. As the
deputy ids seem to be chosen at random with no specific arithmetic
property, there is a one-half probability that they belong to $Q_p$, as
for any element of $\F_p^*$. There could be some bias if the deputy ids
had only a few bits, but with 32-bit integers, according to standard
number theoretic heuristics this will not be the case. A plausible
scenario for the attack is then a district where two candidates
concentrate most of the votes, one of them having a deputy id in $Q_p$
and the other not. Then, from an encrypted ballot, by computing a
Legendre symbol, one can deduce the voter's choice unless she voted for a
less popular candidate.

Therefore, as for the first attack, this second attack means that vote
secrecy relies on a very strong trust assumption in the voting server,
and that the partial results are leaked all along the process.

At first, it seems that the designers were skeptical about the
feasibility of this second attack, and they denied that it was a threat.
However on August 28, 2019, they organized a last public testing, with
only two deputy ids. It would have been fully vulnerable to the described
attack, since one of the ids was in $Q_p$ and the other not. But despite
the public source code was not yet modified, the (minified) Javascript
served to the volunteers during the test included a patch against
our second attack.

\section{Discussion}\label{sec:discuss}

\subsection{The role of the blockchain in the protocol}

\subsubsection{Blockchain as a distributed ledger}

In the protocol, the encrypted ballots are sent to an Ethereum blockchain
and stored as transactions, one transaction per ballot. The argument for
doing so is a typical one used in e-voting, namely offering the
possibility for the voters to check that their vote is indeed taken into
account. At the end of the election, again via the blockchain, the voters
are also given a way to relate each encrypted ballot to the
corresponding vote in cleartext.  The goal is to provide the
cast-as-intended property: if the voting client were to silently modify
the choice of the voter, this would be detected.

In the above quick description, we implicitly assumed that once the voter has done the check that
her ballot is present in the blockchain it will stay there and be counted
in the tally. This also assumes that the voters are given enough
information and tools to record the link between their vote and the
corresponding entry in the blockchain, so that the check can be done in
the few days (and maybe weeks or months) after the election.

In the Moscow election, a specific, permissioned Ethereum blockchain was
used. The impossibility for the nodes running this blockchain to
rewrite the history of the ledger in order to remove a ballot after the voter
has checked it, relies therefore on the assumption that enough nodes are
honest.  Furthermore, the access to this specific blockchain was not
guaranteed to stay for long, and actually was cut by the organizers
quickly after the election.

Without access to the specifications of the protocol it is difficult to
draw strong conclusions, but we consider that the verifiability
properties were not as strong as what could be hoped for from a
blockchain-based ledger.

\subsubsection{To use or not to use a smart contract for decryption}

In the original implementation, at the end of the election, the 3 private
keys of the multilevel ElGamal were used to publicly decrypt all the
ballots. This decryption was implemented in the Solidity programming
language, to be run as part of a smart-contract by the nodes of the
blockchain. The security properties that were sought by doing so are
unclear. There are many ways to guarantee that a decryption has been
correctly done, the most obvious in an ElGamal encryption setting is to
include a simple zero-knowledge proof (as done for instance in
Helios~\cite{HeliosPaper}).

In the version that was modified after our first attack, the protocol was
changed, so that the decryption was done outside the smart-contract. 
The decryption results, namely the votes in clear, were uploaded to the
blockchain as simple transactions with no computation. This operation
occurs of course at the end of the election, in order to compute the
tally. And additionally, the private key was also stored in the
blockchain. This indeed allows the voters to verify that the decryption
is correct.

Doing such a big change in the protocol just a couple of weeks before a
real use in a real and high-stakes election is definitely not a good
practice. However, again, without a proper specification, it is hard to
deduce all the consequences. Did the trust assumptions change in the
process? This also leaves open speculations about the possibility that
programming the decryption in a smart contract was nothing but
a peculiarity of the original design.

\subsubsection{Is it the origin of the small key sizes?}

The original code included many checks ensuring that the primes used to
defined groups for the multilevel ElGamal encryption had a size small enough so
that they would fit in 256 bits. This was taking the form of comparisons to
a constant called {\tt SOLIDITY\_MAX\_INT} defined as $2^{256}-1$. It
indeed corresponds to the largest (unsigned) integer type natively
supported by the Solidity programming language of the Ethereum smart
contracts. A private communication with the designers confirmed that
the reason for removing the ballot decryption of the smart-contract code
and changing the protocol accordingly was due to the lack of time to
implement a multi-precision library in Solidity, that became necessary
after increasing the key size to 1024 bits.

Although the coincidence of the originally chosen bit size for the primes
and the largest integer size natively supported in Solidity is striking,
it is hard to be sure that this is the reason for the mistake. We can
however speculate and consider that the purpose of the multilevel variant
of ElGamal was to compensate for this admittedly small key size. Maybe
the designers hoped for a much better security by using the three
successive encryptions, just like Triple-DES is much stronger than DES.
Unfortunately, things are quite different for asymmetric cryptography.

Another cause of using 256-bit keys could be the confusion between the
security brought by elliptic curve cryptography and the one offered by
using finite fields.

\subsection{What occurred on D-day}

The public source code repository was updated on September 6 (two days
before the election) in order to take into account our second attack. In
the final version the message $m$ to be encrypted is now squared before being passed to the
ElGamal encryption, so that, indeed, the data that is encrypted is a
quadratic residue.

The prime chosen to define the group is congruent to 3 modulo 4. This has
the following consequence: $(-1)$ is a quadratic non-residue in $\F_p^*$,
and the Tonelli-Shanks modular square root algorithm~\cite{GatGer} takes its simplest
form, namely raising to the power $(p+1)/4$.

In order to recover the original message after the decryption, this
square root by modular exponentiation is performed, and the sign choice
is based on the relative size of $p-m$ and $m$ as integers between $1$ and
$p-1$. Indeed, all the deputy identities that are encrypted as integers
that are much smaller than $p$.

This is close to a fix we proposed when publishing our second attack, but
instead of doing an additional exponentiation during encryption and
having a cheap decryption, here the encryption is cheap and the
decryption includes the additional exponentiation. This makes sense,
since the decryption can be done on high-end servers, while the
encryption is done on the voter's device which might be a smartphone.
\medskip

Therefore, on September 8, the election took place with an encryption
procedure which was not easy to break. Even though 1024 bits
are not enough for even a medium-term security, it is certainly hard (not
to say infeasible) to solve a discrete logarithm problem of that size in
less than 12 hours of wall clock time. With the current public
algorithmic knowledge (and extrapolations based on existing record
computations~\cite{logjam,dl768}), billions of computing cores would have to be
mobilized and made to cooperate, which sounds unlikely, even with the
resources of a major company or governmental agency.
\medskip

According to the organizers, more than 10 thousands of Muscovites used
the Internet voting system, in the 3 districts. In one of the districts,
the difference between the first and the second candidates was less
than 100 votes in total. This proves in retrospect the really high stakes
of this experiment, since a risk of fraud in the system directly means a
risk on the final result.

During the election, it was possible to access the blockchain data with a
web interface, and the encrypted ballots were present in it. At the end
of the day, the private key was also sent to the blockchain for
verifiability purposes. But a few hours later, the access to the
blockchain was cut. Fortunately, analysts of the Meduza online newspaper
recorded everything and made the data available%
\footnote{\href{https://meduza.io/slides/meriya-sluchayno-pozvolila-rasshifrovat-golosa-na-vyborah-v-mosgordumu-my-eto-sdelali-i-nashli-koe-chto-strannoe}%
{\nolinkurl{https://meduza.io/slides/meriya-sluchayno...chto-strannoe}}}. They also used the
private key to decrypt the 9810 encrypted ballots they found in the
blockchain and published them. The statistics they observed from this
data raises questions about the fairness of the election, but it is
impossible to draw conclusions from just the published data.

This cutting of the access to the blockchain just after revealing the
decryption key looks like an attempt to mitigate the risk on secret of
the votes, while still having some kind of verifiability. This seems not
to have been convincing: Soon after the election, the head of the Central
Election Commission of the Russian Federation, Ella Pamfilova, made a
public declaration%
\footnote{\url{https://www.kommersant.ru/doc/4095101}}
clearly expressing concern about the results of this experiment and that
in the coming years this should not be extended to the whole territory.

\subsection{On the absence of specification}

In our opinion, the main problems with the Moscow Internet voting system
are:
\begin{itemize}
    \item the absence of a public specification;
    \item the modifications made in a rush, just before the election.
\end{itemize}

In a clear specification, we expect to find much more details about the
task of each entity playing a role in the system. From just the source
code it is not always clear who is supposed to run some part of the code.
What is also needed is clear statements about the security claims and the
trust assumptions.

While the designers obviously had some verifiability properties in mind,
hence used a blockchain, they certainly also wanted to maintain vote
secrecy, as it is always a requirement in such a political context.  It
seems however, that vote secrecy with respect to the web server that
received the (encrypted) ballots was not a goal.  Furthermore, as far as
we can see, coercion-resistance was not at all a concern, at least
initially.

We do not claim that having coercion-resistance and secret with respect
to the
voting server is necessary for any voting system. But this should be
clearly stated, so that the officials who validate the use of the system
can take the decision, while knowing the risks.
\medskip

This ideal process of having a clear specification, with well-stated
trust assumptions and security claims is deeply incompatible with the way
this election was organized. Indeed, while making a slight modification
to a protocol to fix a problem is certainly feasible without having to do
again the security analysis from scratch, the changes made by the
designers just a few weeks or even a few days before the election were so
important that they would have required to revise pages and pages of
documentation if this documentation was public. And in fact, it seems
that the decision to cut the access to the blockchain shortly after the
end of the election was made as a quick response to some bad press about
the risks on privacy and coercion. Somehow, they decided to reduce the
verifiability to try to save other properties.

\section{Lessons learned and conclusion}

The first lesson learned from this story is, not surprisingly, that
designers should be very careful when using cryptography. The authors of
the Moscow system made many mistakes with the encryption scheme they
decided to use. And in fact, even now, technically the encryption is still
weak for two reasons. First, the 1024-bit key is too small for medium
term security, and if the protocol changes so that vote privacy relies on
it, this will not be enough. Furthermore, as far as we could see, the way
the prime was chosen is not public, so that it could include a
trapdoor making discrete logarithms easy to compute for the
designers~\cite{hsnfs}.
Second, textbook ElGamal, which is what is
implemented now, is not IND-CCA2.  Depending on the protocol, this might
lead to minor or devastating attacks. As an example of the latter, in a
protocol that would include a decryption oracle that allows to decrypt
any ciphertext that is not in the ballot box (for instance, for audit
purpose), it would be easy to use the homomorphic properties of ElGamal
to get all the ballots decrypted.

The second lesson is that using a blockchain is not enough to guarantee
full transparency. There are various notions of verifiability in the
e-voting literature~\cite{verifSOK}, and the designers must clearly say which property
they have, under precise trust assumptions. These trust assumptions must
be made even more carefully when using a permissioned blockchain, where
the nodes running the blockchain are probably specifically chosen for the
election, and where the access to the blockchain can be cut at any time.

Even more specific to e-voting, the Moscow system is a good example of
the difficulty for an Internet voting system to make the vote secrecy
rely uniquely on cutting the link between the voters and their encrypted
ballots when they arrive on a server that should also authenticate the
voters. What is really required is to cut the link with the vote in
clear, and, for this, classical methods exist like homomorphic decryption
or verifiable mixnets. In such a high-stakes election, many seemingly
incompatible security properties must be satisfied (secrecy vs
transparency), and advanced cryptographic tools are almost impossible to
avoid.
\smallskip

Finally, as a conclusion, although our attacks led to the
system using a better encryption scheme, it is clear that the system as a
whole is still far from being perfect. We consider it likely that if the
specification were becoming public in the future, other attacks would be
revealed. Therefore, we believe that the main impact
of our work was to draw the attention to the system as something that was
maybe not as secure as what was claimed. The bad publicity in the press
hopefully influenced some potential voters who decided not to take the
risk of using this still really problematic system and went for paper
ballots instead. 

\section{Acknowledgements}
Thanks to Iuliia Krivonosova and Robert Krimmer, for sharing some
information about the Moscow Internet voting. In particular Iuliia's blog
post~\cite{medium} was quite useful. We also thank Noah Stephens-Davidowitz for his comments on an earlier version of this note. We thank Mikhail Zelenskiy and Denis Dmitriev for sharing some data and information about the voting scheme.

\bibliographystyle{plain}
\bibliography{merger}

\newpage
\appendix

\section{A shell script for the first attack}
\label{app:dlog}

{\small
\begin{verbatim}
## These are commands to be run on a Linux machine (Debian or Ubuntu).
## The main tool for the discrete logarithm computations is CADO-NFS,
## and we use GP-Pari as a 'pocket calculator' for modular arithmetic.
# install some packages
sudo apt install pari-gp jq
sudo apt install libgmp3-dev gcc g++ cmake libhwloc-dev
alias gpnoc="gp -q --default colors=\"no\""
# download and compile cado-nfs
cd /tmp
git clone https://scm.gforge.inria.fr/anonscm/git/cado-nfs/cado-nfs.git
cd cado-nfs
git checkout 6b3746a2ec27  # version of 16/08
make cmake
make -j 4

# download blockchain-voting and extract public keys
cd /tmp
git clone https://github.com/moscow-technologies/blockchain-voting.git
cd blockchain-voting
git checkout d70986b2c4da  # most recent version at the time of writing
cd /tmp

# loop on the 3 public keys; could be done in parallel on 3 machines.
for i in {0,1,2}; do
  start=`date +%s`
  # extract the public key information
  keyfile="/tmp/blockchain-voting/encryption-keys/keys/public-key.json"
  p=`jq .modulos[$i] $keyfile | tr -d \"`
  g=`jq .generators[$i] $keyfile | tr -d \"`
  h=`jq .publicKeys[$i] $keyfile | tr -d \"`
  ell=`echo "($p-1)/2" | gpnoc`
  # run cado-nfs to get log of h (takes a few minutes)
  wdir=`mktemp -d /tmp/cadorunXXXXXX`
  log_h=`/tmp/cado-nfs/cado-nfs.py -dlp -ell $ell workdir=$wdir target=$h $p`
  # run again to get log of generator (faster, since it reuses precomputed data)
  log_g=`/tmp/cado-nfs/cado-nfs.py $wdir/p75.parameters_snapshot.0 target=$g`
  # deduce private key
  x=`gpnoc <<EOF
xell=lift(Mod($log_h,$ell)/Mod($log_g,$ell)); half=lift(1/Mod(2,$ell));
x0=lift(Mod(2*half*xell, 2*$ell)); h0=lift(Mod($g,$p)^x0);
if (h0 != $h, x0=lift(Mod(2*half*xell+$ell, 2*$ell)));
x0
EOF`
  stop=`date +%s`
  echo "Private key number $((i+1)) is $x, computed in $((stop-start)) seconds."
done
\end{verbatim}
}

\section{Encrypted messages are not quadratic residues}
\label{app:legendre}

In this Appendix we use the provided public key and encrypted
messages in the modified version of the public repository~\cite{github}
(revision 1d4f348681e9)
to show that not all messages are quadratic residues in $\F_p^*$. Here
$b$ is the set of the second components of the encrypted messages (that
is, each element of the set is $\pk^r\cdot m$ where $m$ is some plain
message, $\pk$ is the public key, and $r$ is a random number). The
following Python code
shows that only five out of ten elements are quadratic residues.
For simplicity, we compute the Legendre symbol by modular exponentiation:
the quadratic residues give 1 when raised to the power $(p-1)/2$.

\begin{lstlisting}
p =
  10062759081450625618037903678618826196600591242500860802791085970455088
  29615914188038720723057459046019130152450978128758867982127126946624453
  23678201384359740027439588690880234391145675099291004487668846511981135
  30933109486902142540395785614572268133031351548262091859360232929939444
  1379077427748866822254003

q = (p-1)//2

b = [
  86911001506497462251782638567319361833688978813664946437333829354738909
  40443974481927929263283486987233406326466505025027434679060583881689706
  23263052860581382950559847777412555501704989450676046755496358356631412
  74356550963994173797345489306417174072514309856175754908122436241421564
  859178326320313204945649,
  32994578715846315625334282465389128113015193084444994471583135772127926
  44951892161427453570566766298979864185170520616403124797427010730707520
  16109483404053598174999416617877699551805519137361275465665467691230764
  44375224889357541488942667685714188203805416972085863674686599803137288
  027861639262227344813980,
  25605451399106620676652873102021964641362454624148409311459772958496440
  67016843578315908545184077772794593830979151616819810966255709567920814
  13077819709806694723689969137957383923170349530451483441188337477065322
  87151838997509598299206147956479381022563215978764100195629663712388182
  647511089787862332483202,
  30936197551567269685847042352240834287171756541862382295858852516666762
  11755805979729879023007285286880732674891989007741022633330800550368742
  56346941237089009381794632389798562078456796442958644789501357076108208
  77962547470703268773776147336174270678101221755152924933175072952910690
  305403946708512011344065,
  90189227659365697355063500941760536836478537551461759945631823319091683
  13130539947043416222984580270526152593756457555485599018740243229324226
  84960561239260442729637671756134870576696053584273031857981168518983390
  53864084929055706240055307151918736952456608210700937953363208336695605
  308414504363789714782355,
  91764714915834445310265717136195446845915020510854708634828807741642908
  65088805234016509342009913809428795919722926613847539079055997816788187
  99170526245002211336442034207826902363786376681934271623388852857592304
  13278401533846260888398253877915981254520562872698617685705979612448346
  470413913994244174120780,
  53180133541691920877303393106622876213880557470163604793597655634027675
  13360685116768376758300338878651961955633191844125587620057500524945640
  23932277996165942274611488630312874402187304375485303772307277867299568
  05232142613661312171461386140429576621530845469410809123204273518058446
  975266361694186911940244,
  96389110287648758509344773386657594488132549702589565012028823522666392
  60323174326871289534690190117827254235251942037419181816826781045590593
  29371155623633657479236340811419693309298082823008055773940379928788914
  61243697630183068655120651685499248763092459000930306871431366198968873
  609555301941599393034947,
  68868518968718401961947565883286957678496859516081208645391394051517430
  60154089569868014396600078685718742310976349636761884312463762214119090
  17014367814111630789237262689248078371187306393398854088463937893954685
  30979657018007065848405280697276892839194542147616119874097494557367533
  44803639667081357573332,
  59049331935932409191703521981449178033897833739363938803374780496048381
  08167852649116009537459386386032599267182731855221804003545963016545542
  41231467392800236514010370577555635998585837533974218865577533874244033
  45003133365685878245562130520111649077186632157205095851334912141011894
  784614717824328145876601
]

for i in range(len(b)):
    print(pow(b[i], q, p))

\end{lstlisting}

\section{Parameters from the public testing held on August 28}
\label{app:testing}

In this Appendix we show the encryption parameters used for the last
public testing held on August 28, 2019. These were not included in the
GitHub public repository but were extracted from the Javascript code sent
to the voters.  There were two candidates, one option corresponded to a
quadratic residue, while the other one corresponded to a quadratic
non-residue. Therefore, the second attack described in
Section~\ref{sec:second_attack} would have decoded all votes.
\begin{lstlisting}
p =
  12270848251665690851841155105748670756648053237913900516699359405362771
  39717263095726449865110213728719981659033550058365258369834144969686617
  29191112587333253191262755602784412922675331893614019119979108938727080
  35007007749458130783976450013979645236359373116042676595576310035726012
  4300619948890487736216143
q = (p-1) // 2
m1 = 3247602110
m2 = 667396531
print(pow(m1, q, p))
print(pow(m2, q, p))

\end{lstlisting}

\end{document}